\documentclass[lettersize,journal]{IEEEtran}
\usepackage{amsmath,amsfonts}
\usepackage{algorithmic}
\usepackage{algorithm}
\usepackage{array}
\usepackage{textcomp}
\usepackage{stfloats}
\usepackage{url}
\usepackage{verbatim}
\usepackage{graphicx}
\usepackage{cite}
\usepackage{placeins}
\usepackage{booktabs}
\usepackage{mathtools}
\usepackage{bbm}
\usepackage{bm}
\usepackage{bbding}
\usepackage{amssymb}
\usepackage[colorlinks=true,linkcolor=blue,citecolor=green,urlcolor=blue,]{hyperref}
\usepackage[caption=false,font=footnotesize,labelfont=rm,textfont=rm]{subfig}
\usepackage{newtxmath}
\usepackage{multirow}
\usepackage[caption=false]{subfig}
\begin{document}

\title{BAP-SRL: Bayesian Adaptive Priority Safe Reinforcement Learning for Vehicle Motion Planning at Mixed Traffic Intersections}

\author{Yuansheng Lian, Ke Zhang, Yaming Guo, Shen Li, Meng Li$^*$

\thanks{This research is supported by grants from National Natural Science Foundation of China (No. 52325209, 52272420), Tsinghua University-Mercedes Benz Joint Institute for Sustainable Mobility, and Tsinghua-Toyota Joint Research Institute Inter-disciplinary Program. \textit{(Corresponding author: Meng Li.)}

Meng Li is with the Department of Civil Engineering, Tsinghua University, Beijing 100084, China, and also with the State Key Laboratory of Intelligent Green Vehicle and Mobility, Tsinghua University, Beijing 100084, China (e-mail: mengli@tsinghua.edu.cn)

Yuansheng Lian and Shen Li are with the Department of Civil Engineering, Tsinghua University, Beijing 100084, China. (e-mail: lys22@mails.tsinghua.edu.cn, sli299@tsinghua.edu.cn). Ke Zhang is with the College of Traffic Management, People's Public Security University of China, Beijing 100038, China. (e-mail: zhangke@ppsuc.edu.cn). Yaming Guo is with the College of Metropolitan Transportation, Beijing University of Technology, Beijing 100124, China, and also with the Department of Civil Engineering, Tsinghua University, Beijing 100084, China (e-mail: guoyaming123@tsinghua.edu.cn).
}}

\markboth{}%
{Shell \MakeLowercase{\textit{et al.}}: A Sample Article Using IEEEtran.cls for IEEE Journals}


\maketitle

\begin{abstract}
Navigating urban intersections, especially when interacting with heterogeneous traffic participants, presents a formidable challenge for autonomous vehicles (AVs). In such environments, safety risks arise simultaneously from multiple sources, each carrying distinct priority levels and sensitivities that necessitate differential protection preferences. While safe reinforcement learning (RL) offers a robust paradigm for constrained decision-making, existing methods typically model safety as a single constraint or employ static, heuristic weighting schemes for multiple constraints. These approaches often fail to address the dynamic nature of multi-source risks, leading to gradient cancellation that hampers learning, and suboptimal trade-offs in critical dilemma zones. To address this, we propose a Bayesian adaptive priority safe reinforcement learning (BAP-SRL) based motion planning framework. Unlike heuristic weighting schemes, BAP formulates constraint prioritization as a probabilistic inference task. By modeling historical optimization difficulty as a Bayesian prior and instantaneous risk evidence as a likelihood, BAP dynamically gates gradient updates using a Bayesian inference mechanism on latent constraint criticality. 
Extensive experiments demonstrate that our approach outperforms state-of-the-art baselines in handling interactions with stochastic, heterogeneous agents, achieving lower collision rates and smoother conflict resolution. 
\end{abstract}



\section{Introduction}
\IEEEPARstart{M}otion planning is the cornerstone of autonomous driving, tasked with generating feasible paths or actions that ensure safety, efficiency, and passenger comfort in dynamic traffic environments \cite{teng2023motion}. 
This task becomes especially complex in urban intersections, where automated vehicles (AVs), human-driven vehicles (HDVs), and vulnerable road users (VRUs) co-exist \cite{geisslinger2023ethical}. The emergent, multi-scale dynamics of such heterogeneous agents introduces profound challenges to autonomous driving safety.

Deep reinforcement learning demonstrates exceptional potential in handling mixed traffic by learning optimal policies through interaction. However, applying standard RL to safety-critical scenarios presents a fundamental structural limitation regarding how safety is incorporated. The predominant approach embeds safety directly into the reward function by assigning large negative penalties to undesirable outcomes, such as collisions \cite{zhang2025carplanner}. While intuitive, this technique conflates the primary objective of task performance with the hard constraints of safety \cite{zhang2021safe}. In complex environments with competing objectives, aggregating safety and performance into a scalar reward often leads to reward hacking or dangerously suboptimal risk-taking behaviors \cite{amodei2016concrete}.

To address these limitations, safe RL has emerged as a robust paradigm. Some safe RL methods apply safety filter or shield to modify existing actions \cite{wang2023safe, wang2024autonomous}. Another category of research formally models the problem as a constrained Markov decision process (CMDP) \cite{garcia2015comprehensive}. Unlike standard RL, these methods explicitly decouple the reward objective from safety constraints during policy optimization. Generally, solutions fall into two categories: primal methods, which enforce constraints directly within the policy search space, and primal-dual methods, which solve a dual optimization problem \cite{ding2020natural}. 

\begin{figure}[t!]
    \centering
    \subfloat[Intersection scenario with multi-objective constraints.]{
        \includegraphics[width=\columnwidth]{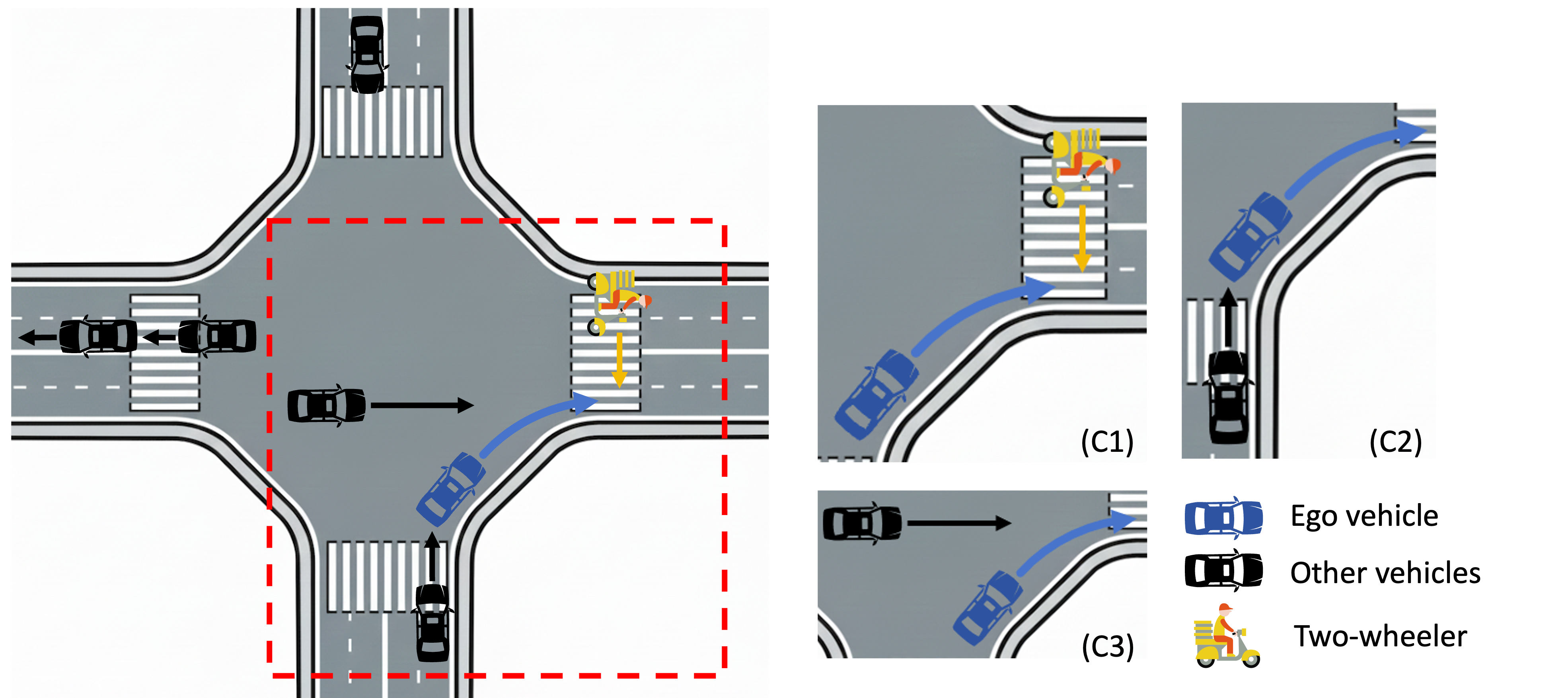}
        \label{fig:scenario_vis}}
        
    \subfloat[Gradient conflict analysis.]{
        \includegraphics[width=0.9\columnwidth]{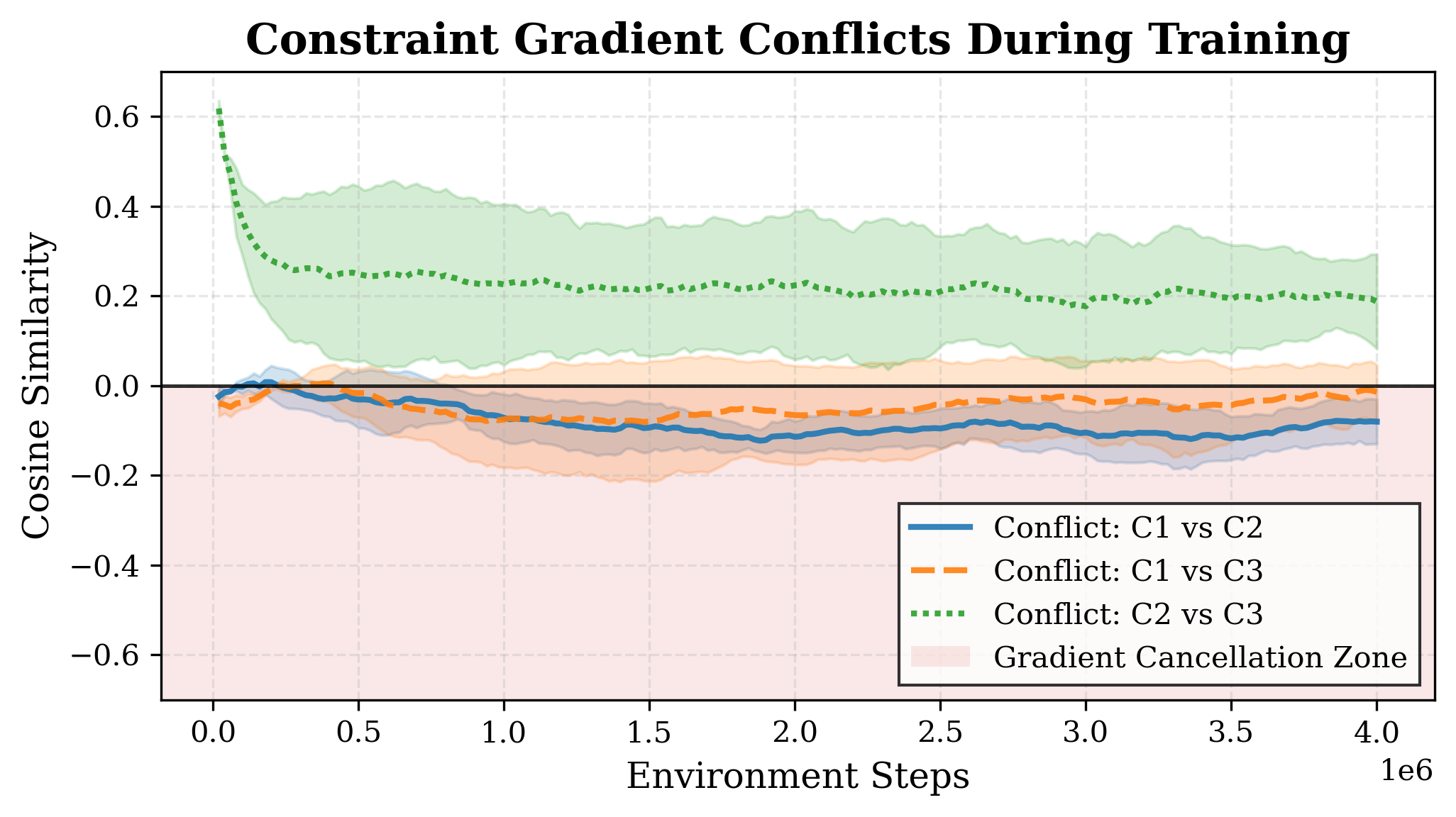}
        \label{fig:conflict_data}}
\caption{\textbf{The challenge of multi-objective conflict in safe autonomous driving.} 
(a) The ego vehicle (blue) must navigate a dynamic intersection facing simultaneous collision threats from heterogeneous traffic agents, creating complex spatiotemporal conflicts. 
(b) An empirical analysis of {constraint gradients} during the training of the PPO-Lagrangian baseline \cite{ray2019benchmarking}. By tracking the cosine similarity between gradients, we observe that some pairs frequently drop into the negative zone. This indicates physical conflicts translate into algorithmic {gradient cancellation}, potentially trapping the safe RL agent in behavioral dilemmas and optimization instability.}
\label{fig:motivation_combined}
\end{figure}

However, realistic urban intersections introduce a higher order of complexity: multi-source traffic risks that impose simultaneous and often conflicting safety constraints \cite{yao2024gradient}. As illustrated in Fig. \ref{fig:scenario_vis}, in a typical mixed-traffic dilemma zone, such as a cyclist cutting in from the front while another vehicle tailgates aggressively from behind or side, the ego vehicle faces opposing safety requirements. Standard primal-dual methods, which typically aggregate constraint gradients using Lagrangian multipliers as static weights, face potential failures in these scenarios. When gradients from opposing constraints conflict, the agent suffers from {gradient cancellation} \cite{yang2022constrained, yu2020gradient}. 
This phenomenon is empirically illustrated in Fig. \ref{fig:conflict_data}, where our analysis of the PPO-Lagrangian baseline \cite{ray2019benchmarking} reveals that the cosine similarity between some constraint gradient pairs frequently drops into the negative zone. This indicates that physical dilemmas translate directly into algorithmic counteractions, leading to policy paralysis or oscillation. Moreover, the episodic evolution of Lagrange multipliers often fails to capture the sub-second risk dynamics inherent in emergent long-tail events \cite{lian2026cdkformer}, leading to delayed safety interventions.


We believe safe RL-based motion planning methods in mixed traffic requires a dynamic, context-aware prioritization mechanism capable of instantaneously rebalancing conflicting constraints. To this end, we propose Bayesian adaptive priority safe reinforcement learning for motion planning at intersections. We conceptualize the importance of each safety constraint not as a fixed hyperparameter, but as a latent random variable. By modeling the system's long-term cost experience as the prior, and the instantaneous cost violation as the likelihood, derived from current violation risks, BAP analytically computes a posterior attention weight for each constraint at every time step. 
This effectively transforms the policy gradient update into a risk-sensitive and adaptive mechanism, thereby mitigating the problem of gradient cancellation. The main contributions of this study are threefold:
\begin{itemize}
    \item We propose a Bayesian adaptive priority safe reinforcement learning framework for autonomous vehicle motion planning in complex mixed traffic intersections. We formulate the safe motion planning with multiple safety constraints as a constrained Markov decision process and employ Lagrangian relaxation to transform the constrained optimization into a dual problem. We treat the importance of different safety constraints as latent random variables, allowing the planner to infer dynamic safety priorities directly from the evolving traffic context.

    \item We propose the Bayesian adaptive priority (BAP) mechanism, a probabilistic gradient shaping technique for multi-objective Lagrangian safe RL. By fusing priority priors from historical experience with likelihoods from instantaneous risk evidence, BAP dynamically gates constraint gradients based on posterior criticality inferred via Bayesian updates. This effectively mitigates the gradient cancellation problem in standard Lagrangian methods, ensuring stable policy optimization in dilemma zones.
    \item We validate our framework in CARLA simulations designed to capture the complexity of mixed traffic environments. Results show that BAP-SRL significantly reduces collision rates without compromising traffic efficiency in safety-critical scenarios involving stochastic, adversarial, and heterogeneous traffic agents, outperforming state-of-the-art baselines.
\end{itemize}

\section{Related Work}

\subsection{RL-Based Motion Planning}
Motion planning at intersections has evolved from rule-based and optimization methods toward learning-based paradigms \cite{zhou2019autonomous, teng2023motion, li2024smpc, lian2025game}. 
In recent years, deep reinforcement learning has emerged as a flexible alternative for navigating dynamic traffic. 
Early applications of RL in motion planning primarily utilized value-based methods, such as deep Q-networks and its extensions \cite{zhang2023predictive}. 
Deshpande et al. \cite{deshpande2020behavioral} propose a deep recurrent Q-network for vehicle motion planning in dense urban scenarios with pedestrian crossing.
He et al. \cite{hu2025perceptual} proposed an Adapt-HD3QN algorithm for motion planning in unprotected left turn scenarios, which incorporates heuristic functions from search-based planning into the reward structure, significantly enhancing learning efficiency compared to standard DQN.

However, value-based methods rely on indirect policy generation via maximization over discrete actions, which is inefficient for high-dimensional vehicle control. To overcome these limitations, researchers have shifted toward actor-critic frameworks, such as proximal policy optimization (PPO) \cite{schulman2017proximal} and soft actor-critic (SAC) \cite{haarnoja2018soft}, which directly optimize policies in continuous action spaces.
Actor-critic framework is able to manage continuous control variables, such as precise steering angles and longitudinal acceleration, which are essential for smooth trajectory generation in mixed traffic scenarios \cite{yang2024cdrp3}. 
Rezaee et al. \cite{rezaee2021motion} utilized distributional RL within SAC frameworks to enhance safety of AV motion planning in occluded pedestrian scenarios. By maximizing the lower bound of value distributions, the method effectively mitigates risk through worst-case assumptions.

In terms of traffic risk management, standard RL approaches predominantly rely on reward shaping, where safety is treated as part of the reward function. As noted by \cite{amodei2016concrete}, incorporating safety into a scalar reward alongside performance leads to reward hacking and fails to provide the rigorous constraint satisfaction required for high-risk driving missions. This necessitates the adoption of safe RL to explicitly consider safety constraints in policy optimization.

\subsection{Safe RL-Based Motion Planning}

Safe RL is a subfield of RL that focuses on developing agents that not only learn to maximize reward, but also adhere to safety constraints \cite{garcia2015comprehensive}.


A wide range of safe RL-based motion planning model aims to modify original unsafe actions with safer alternatives, including pre-defined backup policy (usually emergency braking) \cite{kamran2021minimizing, zhang2022safe, wang2023safe}, control barrier function (CBF)-based approaches \cite{wang2024autonomous, zhang2025safe}, reachability-based approaches \cite{krasowski2022safe, kochdumper2023provably}, learnable safety region \cite{gu2023safe}, etc. These methods offer strong safety guarantees to motion planning, some of which are mathematically verifiable. However, they often rely on prior knowledge of the environments or a large set of situation-dependent rules \cite{muller2022motion}.

Another category of safe RL methods directly performs policy optimization with consideration of safety constraints. This requires the agent to maintain a balance between maximizing task rewards and minimizing constraint violations.
Within this paradigm, primal methods often incorporate safe recovery mechanisms to pull the agent back to the feasible region when constraints are violated \cite{wen2020safe, thananjeyan2021recovery}.
Conversely, Lagrangian-based methods utilize dual variables to transform the constrained problem into an unconstrained one, facilitating flexible trade-offs through gradient-based updates \cite{paternain2022safe, wang2023autonomous}.
Our study falls into this category.

\subsubsection{Multi-Constraint Safe RL}

While standard safe RL methods effectively handle individual safety constraints, urban intersections introduce a higher order of complexity: multi-source traffic risks that impose simultaneous and often conflicting safety requirements \cite{yang2023novel}. 
Most existing planners, however, do not explicitly distinguish among these risks, either considering only a single constraint or treating constraints equally \cite{zhang2024safe}. 
This is problematic in dilemma zones where constraints conflict. As noted by Yang et al. \cite{yang2022constrained}, when multiple constraints generate opposing gradients, the agent suffers from {gradient cancellation}, where the vector sum of updates negates the policy's corrective response, leading to oscillation.

Recent studies have attempted to mitigate these conflicts through specific optimization techniques. 
Zou et al. \cite{zou2022multi} propose a multi-constraint proximal policy optimization method for smooth vehicle control.
Huang et al. \cite{huang2022constrained} consider preference in constrained RL. Preferences are learned by using gradient descent to optimize a fitness function that evaluates whether the current policy's estimated Q-values satisfy the required constraint thresholds.
Alternatively, Yao et al. \cite{yao2024gradient} proposed GradS. Based on Lagrangian approach, GradS excludes redundant gradients and updates the policy by randomly selecting a non-conflicting direction.

These approaches typically rely on static heuristics or certain update rules. When applying to motion planning at intersections, however, they normally fail to capture the {dynamic heterogeneity} of traffic risks. In mixed traffic, safety violations are highly non-stationary and asymmetric. 
Existing strategies struggle to rebalance these priorities during dynamic and critical safety violations.

To bridge this gap, we propose to treat constraint prioritization as a probabilistic inference task. We utilize a Bayesian mechanism to resolve gradient interference by dynamically gating updates based on real-time risk evidence and historical optimization difficulty.


\section{Problem Formulation}

\subsection{Scenario Definition}
As shown in Figure \ref{fig:scenario_vis}, we consider the motion planning problem in a dynamic urban intersection with a set of heterogeneous traffic participants $\mathcal{N} = \{A_1, A_2, \dots, A_n\}$, including the ego vehicle $A_{\text{ego}}$, human-driven vehicles, and various VRUs.
The ego vehicle must navigate through the intersection, potentially passing a series of dilemma zones where safety risks may arise simultaneously from multiple sources. 


\begin{figure*}[htbp]
\centering
{\includegraphics[width=\textwidth]{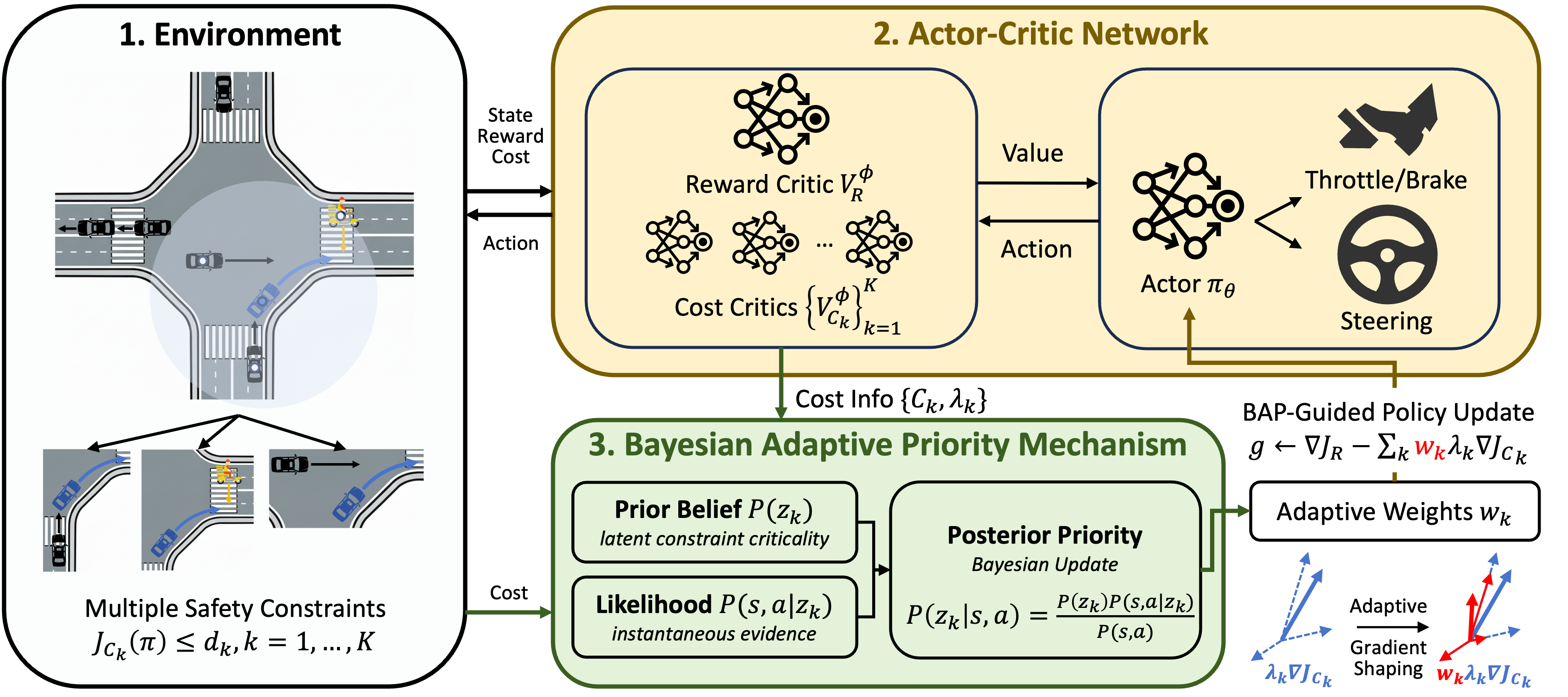}%
\caption{\textbf{The proposed BAP-SRL motion planning framework.}
The architecture is composed of three integrated modules: 
(1) \textbf{Environment}: the ego vehicle perceives the complex intersection environment to form state input while monitoring multiple safety constraints. 
(2) \textbf{Actor-Critic Network}: adopting a multi-critic architecture, the system utilizes a reward critic and independent cost critics to provide value feedback to the actor. The actor generates control actions and updates via a modifed gradient $g$. 
(3) \textbf{BAP Mechanism}: serving as a dynamic safety modulator, this module infers the gradient gating weight in the Lagrangian method by combining the latent prior belief and instantaneous risk evidence via Bayesian updates. The resulting adaptive weights $w_k$ perform sample-wise soft rescaling on cost gradients to dynamically prioritize critical safety violations.
}\label{fig:model-structure}}
\end{figure*}

\subsection{Safe RL Formulation}

In this study, we formalize the vehicle motion planning problem in mixed traffic as a CMDP.
A CMDP extends the Markov decision process by adding a set of cost functions and corresponding constraints. We define it by the tuple $\mathcal{M}_C = (\mathcal{S}, \mathcal{A}, P, R, \{C_k\}_{k=1}^K, \bm{d}, \gamma)$, where:
\begin{align*}
    \mathcal{S}        & : \text{Set of states}                                     \\
    \mathcal{A}        & : \text{Set of actions}                                    \\
    P(s'|s,a)          & : \text{Transition probability}                            \\
    R(s,a)          & : \text{Reward function}                                   \\
    C_k(s,a)        & : \text{The } k\text{-th cost function, for } k=1, \dots, K \\
    \bm{d}             & : \text{Vector of safety thresholds, } \bm{d} = (d_1, \dots, d_K)^T \\
    \gamma             & : \text{Discount factor}
\end{align*}

The objective in a CMDP is to find a policy $\pi$ that maximizes the expected cumulative discounted reward, subject to constraints on the expected cumulative discounted costs:
\begin{align*}
    \max_{\pi} \quad & J(\pi) = \mathbb{E}_{\pi} \left[ \sum_{t=0}^{\infty} \gamma^t R(S_t, A_t) \right] \\
    \text{s.t.} \quad & J_{C_k}(\pi) = \mathbb{E}_{\pi} \left[ \sum_{t=0}^{\infty} \gamma^t C_k(S_t, A_t) \right] \leq d_k \\ & \qquad \qquad \qquad \qquad \qquad \qquad \forall k=1, \dots, K
\end{align*}

Here, $J_{C_k}(\pi)$ represents the expected cumulative discounted cost for the $k$-th cost function. In this study, we consider Lagrangian methods for solving constrained optimization problems. The Lagrangian function $\mathcal{L}(\pi, \bm{\lambda})$ is defined as:
$$ \mathcal{L}(\pi, \bm{\lambda}) = J(\pi) - \sum_{k=1}^K \lambda_k (J_{C_k}(\pi) - d_k) $$
where $\bm{\lambda} = (\lambda_1, \dots, \lambda_K)^T$ are the Lagrange multipliers, with $\lambda_k \geq 0$.
The problem then becomes to solve the minimax problem:
$$ \min_{\bm{\lambda} \geq \bm{0}} \max_{\pi} \mathcal{L}(\pi, \bm{\lambda}) $$
This approach often leads to primal-dual algorithms for solving safe RL problems.
The policy gradient can be calculated as a weighted sum \cite{yao2024gradient}:
\begin{equation}
    \nabla_\theta \mathcal{L}(\pi, \bm{\lambda}) = \nabla_\theta J_R - \sum_{k=1}^K w_k \cdot \lambda_k \nabla_\theta J_{C_k}
    \label{eq:gradient_sum}
\end{equation}
where $w_k$ represents the dynamic weighting coefficient for the $k$-th constraint. Conventional safe RL algorithms \cite{fernando2023mitigating} typically adopt a uniform weighting scheme (i.e., $w_k = 1$), implicitly treating all constraints with equal sensitivity regardless of their instantaneous criticality. The Lagrange multipliers $\lambda_k$ are subsequently updated via dual gradient ascent to enforce the safety thresholds.

\section{BAP-SRL Framework}
\label{sec:methodology}

In this section, we present the BAP-SRL framework for autonomous vehicle motion planning at mixed traffic intersections. 
We formulate the motion planning task as a CMDP and propose a novel probabilistic inference mechanism to resolve weights for conflicting safety constraints.

\subsection{Framework Overview}

The overall architecture of the proposed BAP-SRL framework is illustrated in Fig. \ref{fig:model-structure}. 
Fundamentally, our method operates within the continuous actor-critic paradigm \cite{schulman2017proximal}, extending the standard PPO-Lagrangian formulation \cite{ray2019benchmarking} to address multi-objective safety conflicts. 
The planning process initiates with a {vectorized scene description}, which encodes the ego vehicle's kinematics and the semantic topology of the complex intersection environment into a compact state representation. 
This state input feeds into a dual-branch network structure: a policy network that outputs continuous control actions (throttle/brake and steering), and a multi-head value network  comprising one reward critic and $K$ independent cost critics.

Distinct from traditional methods that simply aggregate costs, our framework integrates the BAP module during the policy update phase. By synthesizing the latent prior belief of constraint difficulty with the instantaneous risk evidence provided by the cost critics, the module computes adaptive gating weights $w_k$. These weights adaptively shape the policy gradient, forcing the planner to prioritize critical safety violations in real-time while maintaining efficiency.
Crucially, the BAP mechanism operates exclusively as an on-policy gradient modulator during the training phase. At inference, the BAP module is detached and the ego vehicle relies on the optimized policy network to execute maneuvers, ensuring real-time performance without additional computational overhead.

\subsection{Environment}
\label{sec:state_network}

As illustrated in Modules 1 of Fig. \ref{fig:model-structure}, our framework relies on a vectorized state representation to encode the complex intersection environment, which serves as the input for the actor-critic network.

To enable comprehensive situational awareness in mixed traffic, we construct a composite state vector $s_t$ defined in the Frenet coordinate system. This vector fuses ego-kinematics, environmental topology, and the dynamic states of surrounding heterogeneous agents. 
Let the ego vehicle be denoted as $A_0$ and the set of observable surrounding agents within the detection range $d_{\rm obs}$ be $\mathcal{N} = \{A_1, ..., A_n\}$. 
The state $s_t$ is formulated as a concatenation of three feature groups:

\begin{itemize}
    \item Ego motion state $s_{\rm ego}$: Encodes the physical status of the ego vehicle relative to the reference path. It includes the lateral deviation $d$, longitudinal velocity $\dot{s}$, lateral velocity $\dot{d}$, and relative heading error $\psi_{\rm rel}$.

    \item Social interaction state ($s_{\rm soc}$): To capture multi-agent interaction dynamics, we track heterogeneous agents relative to the ego vehicle's Frenet frame. For each targeted agent $A_i \in \mathcal{N}$, we extract a feature vector $s_{\rm soc}^i = [\mathbb{I}_{\rm exist}, \mathbb{C}_{\rm type}, \Delta s, \Delta d, \Delta \dot{s}, \Delta \dot{d}]$. Here, $\mathbb{I}_{\rm exist}$ is a binary visibility flag, $\mathbb{C}_{\rm type}$ encodes the semantic category (e.g., pedestrian, two-wheeler, or vehicle), and $(\Delta s, \Delta d)$ and $(\Delta \dot{s}, \Delta \dot{d})$ denote the relative longitudinal/lateral displacements and velocities, respectively.
    
    \item Topology state $s_{\rm map}$: Provides the semantic road context necessary for path tracking. It consists of the relative longitudinal distance and heading error at $K_{\rm wp}$ lookahead waypoints sampled along the reference lane centerline.
\end{itemize}

Finally, the state vector $s_t$ is constructed by concatenating these feature groups:
\begin{equation}
    s_t = [s_{\rm ego}, s_{\rm soc}, s_{\rm map}]
\end{equation}

\subsection{Actor-Critic Network}
\label{sec:network_arch}
The BAP-SRL framework relies on an actor-critic architecture to support its probabilistic inference mechanism, illustrated by Module 2 in Fig. \ref{fig:model-structure}.
The raw state vector $s$ is first processed through a shared multi-layer perceptron (MLP) encoder to extract high-level latent features. To explicitly address multiple distinct safety constraints, the framework employs a policy network alongside a multi-head value network architecture.

The policy network $\pi_\theta$ parameterizes a diagonal Gaussian distribution $\mathcal{N}(\mu_\theta(s), \Sigma)$. The mean $\mu_\theta(s)$ is output by an MLP with a linear final layer, while the standard deviation $\Sigma$ is treated as a state-independent learnable parameter, optimized jointly with the network weights via gradient descent.
The stochastic action $\tilde{a}_t$ is sampled and clipped to adhere to the normalized range:
\begin{equation}
    \tilde{a}_t \sim \mathcal{N}(\mu_\theta(s), \Sigma), \quad a_t = \text{clip}(\tilde{a}_t, -1, 1)
\end{equation}
where $a_t$ represents the normalized control vector $[a_{\mathrm{lon}}, \delta_{\mathrm{steer}}]^\top$.


Unlike standard approaches that aggregate various costs into a single scalar, our design maintains separate evaluation branches implemented via standard MLPs, comprising one reward critic $V^{\phi}_R(s)$ to estimate the expected task return and $K$ independent cost critics $\{V^{\phi}_{C_k}(s)\}_{k=1}^K$ corresponding to different constraint types. These decoupled cost critics provide the detailed risk assessment signals required by the subsequent BAP mechanism.

\subsection{Bayesian Adaptive Priority Mechanism}
\label{sec:BAP_mechanism}

\begin{figure}[t]
    \centering
    \includegraphics[width=\linewidth]{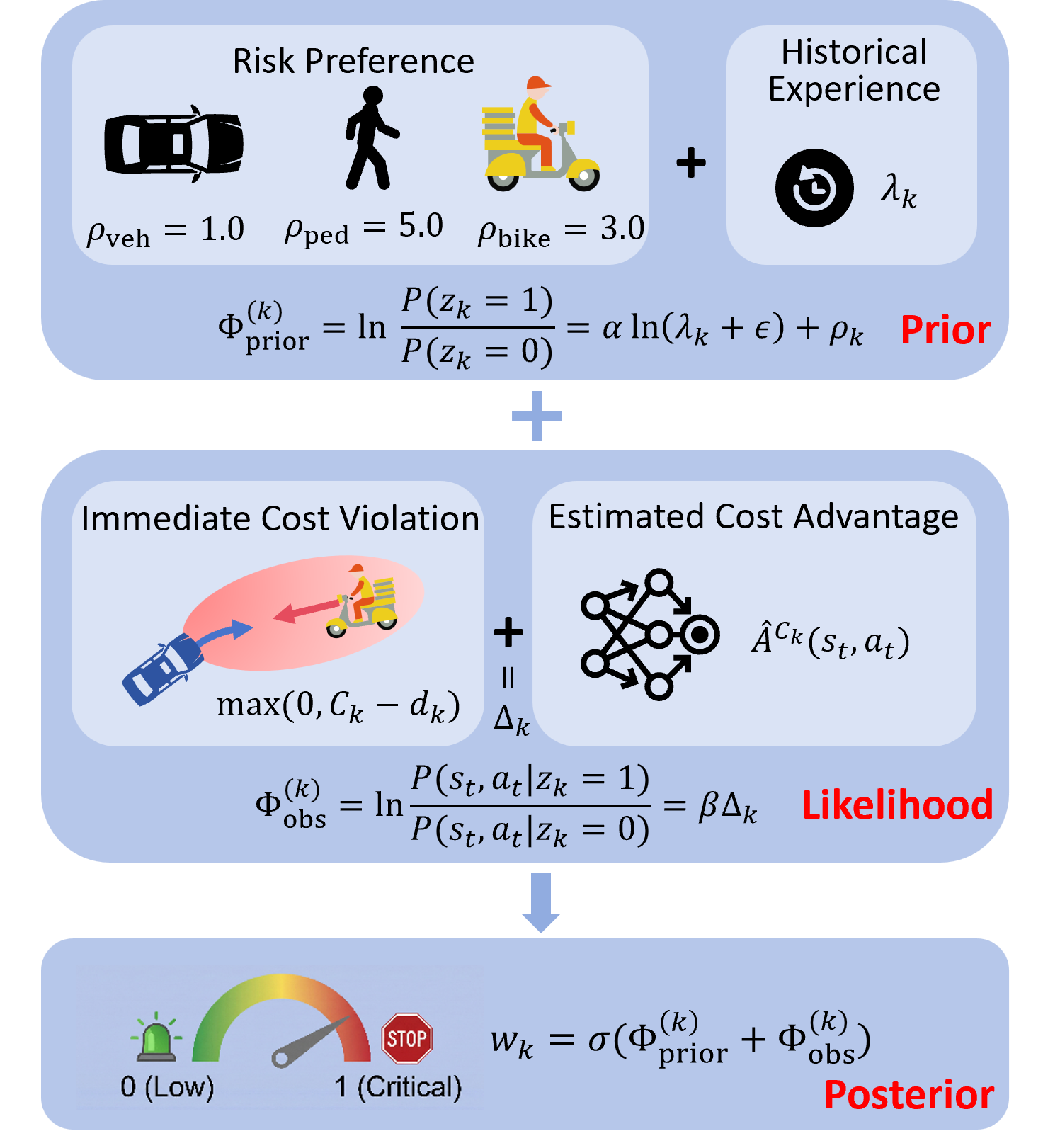} 
    \caption{\textbf{Illustration of the Bayesian adaptive priority mechanism.} The framework models constraint prioritization as a probabilistic inference task, fusing {prior belief} (derived from static priority preferences and historical difficulty ) with {likelihood evidence} (combining immediate violation magnitude and predictive cost advantage). The final posterior weight is computed by Bayesian rule to dynamically gate the policy gradient.}
    \label{fig:bap_mechanism}
\end{figure}

Standard Lagrangian relaxation methods typically treat safety constraints with uniform importance \cite{fernando2023mitigating}, or rely on heuristic balancing techniques that struggle to adapt to highly dynamic environments \cite{yao2024gradient}. However, motion planning in mixed traffic introduces a fundamental multi-objective conflict driven by the heterogeneity of traffic participants. This raises two main challenges. First, safety risks exhibit an inherent asymmetry. Collisions with VRUs carry significantly higher ethical and physical severity than vehicle-vehicle interactions, necessitating differentiated prioritization in the optimization process \cite{geisslinger2023ethical}. Second, a large and emergent safety violation requires a rapid increase in constraint weighting to force an immediate policy correction, whereas minor deviations can be resolved more gradually.  Consequently, an effective safe motion planner requires a mechanism that can jointly integrate historical priority preference with dynamic risk signals.

To address these challenges, we propose the Bayesian adaptive priority mechanism, as shown in Fig. \ref{fig:bap_mechanism}. The core intuition is to mimic the human cognitive process of risk assessment, which fuses long-term experience with immediate risk perception \cite{knill2004bayesian, kolekar2020human}. We formalize this fusion as a probabilistic inference task. We treat the criticality of each safety constraint $k$ as a latent binary random variable $z_k \in \{0, 1\}$, where $z_k=1$ indicates that the $k$-th constraint is currently a critical bottleneck. Our goal is to infer the posterior probability:
\begin{align}
    w_k(s_t, a_t) = P(z_k=1 | s_t, a_t)
\end{align}
where $s_t$ is the current state, $a_t$ is the current action, and $w_k$ serves as a dynamic attention weight in Eq. \ref{eq:gradient_sum} for policy update.

\subsubsection{Probabilistic Formulation}
Starting from Bayesian Theorem, the posterior probability of constraint criticality is derived as:
\begin{equation}
    P(z_k=1 | s_t, a_t) = \frac{P(s_t, a_t | z_k=1) \cdot P(z_k=1)}{P(s_t, a_t)}
\end{equation}
Directly computing the marginal likelihood $P(s_t, a_t)$ is intractable. However, by considering the {odds ratio} of the constraint being critical ($z_k=1$) versus non-critical ($z_k=0$), we can eliminate the marginal term:
\begin{equation}
    \frac{P(z_k=1 | s_t, a_t)}{P(z_k=0 | s_t, a_t)} = \underbrace{\frac{P(s_t, a_t | z_k=1)}{P(s_t, a_t | z_k=0)}}_{\text{Likelihood Ratio}} \cdot \underbrace{\frac{P(z_k=1)}{P(z_k=0)}}_{\text{Prior Odds}}
\end{equation}
This formulation naturally decouples the contribution of instantaneous evidence and historical belief. To facilitate numerical stability and integration with gradient-based optimization, we map this ratio to the logarithmic domain. The final attention weight $w_k(s_t, a_t)$ is obtained by applying the sigmoid function to the sum of log-odds components:
\begin{equation}
    w_k(s_t, a_t) = \sigma \left( \Phi_{\text{obs}}^{(k)}(s_t, a_t) + \Phi_{\text{prior}}^{(k)} \right) \label{eq:w_k}
\end{equation}
where $\Phi_{\text{obs}}^{(k)} = \ln \frac{P(s_t, a_t | z_k=1)}{P(s_t, a_t | z_k=0)}$ represents the evidence from current observation, and $\Phi_{\text{prior}}^{(k)} = \ln \frac{P(z_k=1)}{P(z_k=0)}$ represents the prior belief derived from training history. $\sigma(x) = (1 + e^{-x})^{-1}$ is the sigmoid function.

\subsubsection{Prior Formulation}
The prior belief $P(z_k=1)$ establishes the agent's baseline sensitivity toward constraint $k$ before accounting for instantaneous state observations. In our framework, this prior combines both historical optimization experience and static risk preference.


In the primal-dual framework, the Lagrange multiplier $\lambda_k$ serves as an indicator of constraint satisfaction difficulty. A large $\lambda_k$ implies that the agent has historically struggled to satisfy the $k$-th constraint, necessitating a  repulsive gradient. However, $\lambda_k$ often exhibits high dynamic range variations during training. To mitigate numerical instability, we model the prior odds using logarithmic scaling, allowing the agent to remain sensitive to relative changes in difficulty.

While this adaptive mechanism effectively tracks historical violation frequency, it functions as a purely data-driven feedback lacking {semantic awareness} of the traffic environment. In mixed traffic, risks exhibit inherent asymmetry. For instance, a collision with a high speed two-wheeler is distinct from being rear-ended by a slow moving vehicle. To encode this domain knowledge, we introduce a static bias $\rho_k$ representing the priority of the constraint. This ensures that safety-critical constraints maintain a higher base priority even when their corresponding multipliers are low.

Combining these two components, we formulate the prior log-odds as:
\begin{equation}
    \Phi_{\rm prior}^{(k)} = \ln \frac{P(z_k=1)}{P(z_k=0)} \triangleq {\alpha \cdot \ln(\lambda_k + \epsilon)} + {\rho_k} \label{eq:prior}
\end{equation}
where $\alpha > 0$ is a temperature parameter controlling the sensitivity to historical feedback, and $\epsilon$ is a small constant for numerical stability. $\rho_k$ is the static priority weights relavant to the agent type.

\subsubsection{Likelihood Formulation}
The likelihood term $P(s_t, a_t | z_k)$ captures the instantaneous evidence supporting the hypothesis that constraint $k$ is critical. Unlike standard approaches that rely solely on current constraint violation, we argue that the risk evidence must be both {reactive} to immediate violations and {predictive} of future risks.

To achieve this, we define the violation $\Delta_k(s_t, a_t)$ as the combination of the immediate cost violation with the estimated cost advantage:
\begin{equation}
    \Delta_k(s_t, a_t) = \eta \cdot {\max(0, C_k(s_t, a_t) - d_k)} +  {\hat{A}^{C_k}(s_t, a_t)}
    \label{eq:violation}
\end{equation}
where $C_k(s_t, a_t)$ denotes the immediate cost provided by the current state-action pair $(s_t, a_t)$. 
$\hat{A}^{C_k}$ is the generalized advantage estimation (GAE) \cite{schulman2015high, schulman2017proximal} for the $k$-th cost. It quantifies the expected excess risk of taking action $a_t$ compared to the average behavior of the current policy. A positive $\hat{A}^{C_k}$ indicates that the current action $a_t$ is likely to lead to higher costs than expected. $\eta$ is a scaling factor balancing the two terms. Then we model the likelihood ratio as follows:
\begin{equation}
    \Phi_{\text{obs}}^{(k)} = \ln \frac{P(s_t, a_t | z_k=1)}{P(s_t, a_t | z_k=0)} \triangleq \beta \cdot \Delta_k(s_t, a_t) \label{eq:likelihood}
\end{equation}
where $\beta$ is a sensitivity coefficient.

Finally, the BAP mechanism acts as a differentiable gate for the Lagrangian relaxation. The BAP-based policy gradient update is:
\begin{equation}
    \nabla_{\theta} \mathcal{J}_{BAP} = \nabla_{\theta} J_R - \sum_{k=1}^{K} w_k(s_t, a_t) \cdot \lambda_k \cdot \nabla_{\theta} J_{C_k}(s_t, a_t)
\end{equation}

The complete training procedure, integrating the Bayesian inference loop with the primal-dual optimization, is summarized in Algorithm \ref{alg:bap_srl}.

\begin{algorithm}[t]
\caption{Bayesian Adaptive Priority Safe RL)}
\label{alg:bap_srl}
\begin{algorithmic}[1]
\REQUIRE
    Initialized Actor $\pi_\theta$ and Critics $\{V^\phi_R, \{V^\phi_{C_k}\}_{k=1}^K\}$ \\
    Initialized Lagrangian multipliers $\bm{\lambda} \in \mathbb{R}^K_{\ge 0}$, static priors $\bm{\rho}$ \\
    Hyperparameters: prior temp $\alpha$, likelihood sensitivity $\beta$, step sizes $\eta, \alpha_\lambda$

\FOR{epoch $m = 1, \dots, M$}
    \item[] \textbf{\textit{Phase 1: Interaction \& Data Collection}} 
    \STATE Collect trajectory batch $\mathcal{D}_m$ by executing $\pi_\theta$ in environment
    \STATE Compute Generalized Advantage Estimations $\hat{A}^R_t$ and $\{\hat{A}^{C_k}_t\}$ for all steps
    
    \item[] \textbf{\textit{Phase 2: Bayesian Priority Inference}}
    \FOR{each timestep $t$ in $\mathcal{D}_m$}
        \FOR{each constraint $k = 1, \dots, K$}
            \STATE Calculate prior log-odds $\Phi_{\text{prior}}^{(k)}$ via Eq. (\ref{eq:prior}) using current $\lambda_k, \rho_k$
            \STATE Calculate likelihood ratio $\Phi_{\text{obs}}^{(k)}$ via Eq. (\ref{eq:likelihood})
            \STATE Infer posterior attention weight $w_{k,t}$ via Eq. (\ref{eq:w_k})
        \ENDFOR
        \STATE Calculate advantage $\hat{A}_t^{\text{BAP}}$ via Eq. (\ref{eq:adv})
    \ENDFOR
    
    \item[] \textbf{\textit{Phase 3: Primal-Dual Optimization}}
    \STATE Update policy $\theta$ to maximize PPO objective using $\hat{A}_t^{\text{BAP}}$
    \STATE Update critics $\phi$ to minimize MSE
    \FOR{each constraint $k = 1, \dots, K$}
        \STATE Update Lagrange multiplier $\lambda_k$ via dual ascent:
        \STATE \quad $\lambda_k \leftarrow \max\left(0, \lambda_k + \alpha_\lambda (J_{C_k}(\mathcal{D}_m) - d_k)\right)$
    \ENDFOR
\ENDFOR
\end{algorithmic}
\end{algorithm}

\subsection{Learning Objectives}
\label{subsec:loss_functions}

The framework jointly optimizes the policy parameters $\theta$, value network parameters $\phi$, and Lagrangian multipliers $\bm{\lambda}$. We adopt PPO \cite{schulman2017proximal} as the base algorithm. 
We replace the standard advantage in the PPO surrogate loss with a BAP-modified advantage function to integrate adaptive safety priorities. 
First, we compute the raw GAE, denoted as $\hat{A}^R_t$ for the reward and $\{\hat{A}^{C_k}_t\}_{k=1}^K$ for each safety cost. These advantages are then synthesized into a unified function $\hat{A}_t^{\text{BAP}}$:
\begin{equation}
    \hat{A}_t^{\text{BAP}} = \frac{1}{1 + \sum_{j} \lambda_j} \left( \hat{A}^R_t - \sum_{k=1}^K w_k(s_t, a_t) \lambda_k \hat{A}^{C_k}_t \right) \label{eq:adv}
\end{equation}
This formulated advantage acts as a adaptive gradient weight. It penalizes actions that carry high predictive risk specifically when the corresponding constraint is critical ($w_k \rightarrow 1$) and the Lagrange multiplier $\lambda_k$ is high, while the normalization factor ensures numerical stability.

Consequently, the policy $\pi_\theta$ is updated by minimizing the clipped surrogate objective utilizing this modified advantage:
\begin{equation}
    \mathcal{L}_{\pi}(\theta) = - \mathbb{E}_t \left[ \min \left( r_t(\theta) \hat{A}_t^{\text{BAP}}, \text{clip}(r_t(\theta), 1-\epsilon, 1+\epsilon) \hat{A}_t^{\text{BAP}} \right) \right]
\end{equation}
where $r_t(\theta) = \frac{\pi_\theta(a_t|s_t)}{\pi_{\theta_{\text{old}}}(a_t|s_t)}$ denotes the probability ratio, and $\epsilon$ is the clipping hyperparameter. Simultaneously, the value networks are trained to minimize the mean squared error (MSE) between the estimated values and their respective targets ($y^R_t$ and $y^{C_k}_t$). The joint loss function for the reward critic and the multi-head cost critics is defined as:
\begin{equation}
    \mathcal{L}_{V}(\phi) = \mathbb{E}_t \left[ \left( V^{\phi}_R(s_t) - y^R_t \right)^2 + \frac{1}{K} \sum_{k=1}^K \left( V^{\phi}_{C_k}(s_t) - y^{C_k}_t \right)^2 \right]
\end{equation}
Finally, the Lagrange multipliers are updated via dual ascent to enforce the safety thresholds $d_k$:
\begin{equation}
    \lambda_k \leftarrow \max\left(0, \lambda_k + \alpha_\lambda (J_{C_k}(\pi) - d_k)\right)
\end{equation}
where $J_{C_k}$ represents the expected episodic cost and $\alpha_\lambda$ is the step size.

\section{Experimental Settings}

\subsection{Simulation Environment}
We conduct all experiments in CARLA 0.9.15, utilizing one signalized four-way intersection in Town 03 as the primary testing ground. The simulation runs at a frequency of 20 Hz on a Ubuntu 20.04 workstation equipped with an NVIDIA GeForce RTX 4060 Ti GPU.

The ego vehicle is tasked with navigating the intersection to reach a randomly assigned destination, requiring the execution of three maneuvers: proceeding straight, turning left, or turning right. Background traffic is managed by the CARLA traffic manager.

\subsection{Scenario Design} \label{sec:asd}
To evaluate the proposed framework under safety-critical conditions, we construct an adversarial scenario involving heterogeneous traffic agents. The ego vehicle is simultaneously challenged by three key actors: a red-light running cyclist, a rear vehicle, and a side vehicle. 


The behavior of the VRU is governed by a finite state machine. The VRU remains stationary until the ego vehicle approaches within a threshold distance ($\epsilon_d=25$~m). Upon activation, the VRU samples an intention from a categorical distribution: 
\begin{equation}
    I \sim \text{Categorical}(\phi_{\text{rush}}=0.4, \phi_{\text{yield}}=0.3, \phi_{\text{hesitate}}=0.3)
\end{equation}
In the {rush mode}, the VRU rushes through the intersection at a high speed of {$v_{\text{rush}} \sim \mathcal{U}(5, 7)$~m/s}. The {hesitant mode} mimics {deceptive} yielding behavior by initially decelerating, then accelerating to $v_{\text{rush}}$ after a stochastic pause $\Delta t_{\text{pause}} \sim \mathcal{U}(0.5, 1.5)$~s. $\mathcal{U}(a, b)$ denotes a uniform distribution defined on the interval $[a,b]$.

We also introduce adversarial vehicles to restrict both longitudinal and lateral maneuvering space. A rear vehicle is configured to maintain a narrow following gap of $d_{\text{rear}} = 1$~m, significantly constraining the ego vehicle's emergency braking capabilities. Simultaneously, a side vehicle, generated alongside and traveling in the same direction, operates with a free lane-changing policy near the ego vehicle. 

\subsection{RL Settings}
\label{subsec:rl_settings}

 The definitions of the reward function and safety costs are detailed below.



\subsubsection{Reward Function}
The reward function is designed to balance efficiency, safety, and path tracking accuracy. It is formulated as a weighted sum of components:
\begin{align}
    R &= R_{\rm efficiency} + R_{\rm track} + R_{\rm terminal} + R_{\rm risk}\\
    R_{\rm efficiency} &= w_v \cdot \max\left(0, 1 - \frac{|v - v_{\rm tgt}|}{v_{\rm tgt}}\right) - w_{\rm idle} \cdot \mathbb{I}(v < v_{\rm idle}) \nonumber \\ 
    R_{\rm track} &= -w_{\rm track} \cdot d^2 \nonumber \\
    R_{\rm terminal} &= w_{\rm goal} \cdot \mathbb{I}(\text{goal}) - w_{\rm collision} \cdot \mathbb{I}(\text{collision}) \nonumber \\
    R_{\rm risk} &= - w_{\rm risk}\sum_{i \in \mathcal{N}} P_i H_i \nonumber
\end{align}
where $w_v, w_{\rm idle}, w_{\rm track}, w_{\rm goal}, w_{\rm collision}$ and $w_{\rm risk}$ are positive weighting coefficients. $R_{\rm efficiency}$ encourages maintaining a target velocity $v_{\rm tgt}$ and penalizes velocities below the idling threshold $v_{\rm idle}$. The tracking reward $R_{\rm track}$ penalizes the squared lateral deviation $d$ from the reference path. $R_{\rm terminal}$ provides a sparse reward for reaching the goal and a penalty for collisions.

The risk reward $R_{\rm risk}$ aggregates risks from the set of surrounding agents $\mathcal{N}$. Following \cite{geisslinger2023ethical}, it is defined as the product of collision probability $P_i$ and potential harm $H_i$. We define potential harm $H_i$ as the velocity change ($\Delta v$) experienced by the ego vehicle:
\begin{equation}
    H_i = \Delta v_{\text{ego}} = \frac{m_i}{m_{\text{ego}} + m_i} \sqrt{v_{\text{ego}}^2 + v_i^2 - 2v_{\text{ego}} v_i \cos \alpha_i}
    \label{eq:delta_v}
\end{equation}
where $m_{\text{ego}}$ and $m_i$ are the masses of the ego vehicle and agent $i$, $v_{\text{ego}}$ and $v_i$ are their velocities, and $\alpha_i$ is the angle between their velocity vectors. 

We utilize time-to-collision (TTC) for collision probability estimation. The probability $P_i$ is modeled as a continuous function that increases non-linearly as TTC approaches zero:
\begin{equation}
    P_i(s) = \begin{cases} 
        \left(1 - \frac{TTC_i}{\tau_{\text{TTC}}}\right)^2 & \text{if } 0 \le TTC_i \le \tau_{\text{TTC}} \\
        0 & \text{if } TTC_i > \tau_{\text{TTC}}
    \end{cases}
    \label{eq:prob_ttc}
\end{equation}
where $\tau_{\text{TTC}}$ is a predefined safety time threshold.

\subsubsection{Cost Function}

We employ a hybrid cost formulation that integrates sparse and dense costs.

\begin{equation}
    \mathbf{C}(s, a) = \left[ C_{\text{sparse}}^{(1)}, \dots, C_{\text{sparse}}^{(N)}, C_{\text{dense}}^{(1)}, \dots, C_{\text{dense}}^{(N)} \right]^\top
\end{equation}
where $N$ denotes the number of interacting agents.

We define sparse constraints to strictly prohibit catastrophic failures. For any agent $i$, a binary cost is incurred upon impact:
\begin{equation}
    C_{\text{sparse}}^{(i)}(s) = c_{\rm sparse} \cdot \mathbb{I}(\text{collision with agent } i)
\end{equation}
To provide informative gradient signals before a collision occurs, we incorporate dense cost functions derived from the risk metrics defined in the reward section. Specifically, the dense cost corresponds to the magnitude of the potential safety violation:
\begin{equation}
    C_{\text{dense}}^{(i)}(s) =  c_{\rm dense} \cdot P_i(s) H_i(s)
\end{equation}
This formulation ensures that the agent is penalized not only for actual collisions but also for entering high-risk states.



\subsection{Hyperparameters}
Our BAP-SRL framework is implemented on top of the OmniSafe library \cite{ji2024omnisafe}. The policy and value networks are modeled using multi-layer perceptrons (MLPs). Detailed hyperparameters for PPO training, Lagrangian relaxation, and the BAP mechanism are summarized in Table \ref{tab:hyperparameters}.

\begin{table}[t!]
\centering
\small
\caption{Hyperparameter Settings for BAP-SRL Training}
\label{tab:hyperparameters}
\renewcommand{\arraystretch}{1.15}
\begin{tabular}{lc}
\toprule
{Parameter} & {Value} \\
\midrule
\multicolumn{2}{l}{\textit{\textbf{BAP Mechanism}}} \\
Prior temperature $\alpha$ & 1.0 \\
Likelihood sensitivity $\beta$ & 3.0 \\
Static priority $\rho_{\rm VRU}, \rho_{\rm veh,side}, \rho_{\rm veh,rear}$ & $0.0, -1.5, -2.0$ \\
Advantage scaling factor $\eta$ & 0.01 \\
\midrule
\multicolumn{2}{l}{\textit{\textbf{Lagrangian \& Safety Constraints}}} \\
Lagrangian learning rate $\alpha_{\lambda}$ & 0.035 \\
Initial multiplier $\lambda_{\rm init}$ & 0.001 \\
Cost weights $c_{\rm dense}, c_{\rm sparse}$ & $5.0, 50.0$ \\
Cost limits $d_k$ & $[0.1, 0.1, 0.1, 100, 20, 20]$ \\
TTC threshold $\tau_{\rm TTC}$ & 1.5 \\
\midrule
\multicolumn{2}{l}{\textit{\textbf{Reward Function Weights}}} \\
Velocity tracking $w_v, w_{\rm idle}$ & 3.0, 0.5 \\
Path tracking $w_{\rm track}$ & 0.1 \\
Terminal states $w_{\rm goal}, w_{\rm collision}$ & 100.0, 100.0 \\
Risk penalty $w_{\rm risk}$ & 5.0 \\
\midrule
\multicolumn{2}{l}{\textit{\textbf{PPO Optimization \& Network}}} \\
Total steps $T_{\rm total}$ & $4 \times 10^6$ \\
Steps per epoch & $2 \times 10^5$ \\
Hidden layer sizes & 128 \\
\bottomrule
\end{tabular}
\end{table}

\subsection{Compared Methods}

To comprehensively evaluate the effectiveness of the proposed BAP-SRL framework, we benchmark it against diverse algorithms:
\begin{itemize}
    \item Unconstrained: TRPO \cite{schulman2015trust}, PPO \cite{schulman2017proximal}
    \item Constrained: CPO \cite{achiam2017constrained}, CUP \cite{yang2022constrained}, PPO-Lagrangian \cite{ray2019benchmarking}
\end{itemize}
Since our core contribution lies in resolving conflicts between multiple safety constraints, we also compare BAP-SRL against PPO-Lagrangian equipped with varying constraint integration strategies: vanilla \cite{ray2019benchmarking}, CRPO \cite{xu2021crpo}, Min-Max and GradS \cite{yao2024gradient}.


\subsection{Evaluation Metrics}
\label{sec:metrics}


To assess the performance of the proposed framework, we report the mean and standard deviation of the following metrics over $N_{\rm eval}=100$ evaluation episodes.

\subsubsection{Safety Metrics}
Safety is the primary objective of our method. We evaluate it using two complementary indicators:
\begin{itemize}
    \item {Collision rate (CR):} Defined as the percentage of episodes that terminate due to a collision:
    \begin{equation}
        \text{CR} = \frac{1}{N_{\rm eval}} \sum_{i=1}^{N_{\rm eval}} \mathbb{I}(\text{collision}_i) \times 100\%
    \end{equation}
    where $\mathbb{I}(\cdot)$ is the indicator function.
    
    \item {Average risk:} While CR measures catastrophic failures, average risk captures the agent's exposure to danger during operation. It is calculated as the mean value of the total dense cost accumulated over all time steps.
\end{itemize}

\subsubsection{Efficiency Metrics}
\begin{itemize}
    \item {Average speed:} The mean velocity maintained by the ego vehicle throughout the episode.
    \item {Time to goal (TTG):} The average duration required to complete the navigation task. Note that TTG is computed only for successful episodes.
\end{itemize}

\subsubsection{Comfort Metrics}
Ride comfort is evaluated using {average jerk}, defined as the mean magnitude of the rate of change of acceleration over the episode duration $T$:
\begin{equation}
    \text{Jerk} = \frac{1}{T} \int_{0}^{T} \|\dot{\mathbf{a}}_t\|_2 \, dt
\end{equation}

\section{Results}
\label{sec:results}

\begin{figure*}[t]
    \centering
{\includegraphics[width=\textwidth]{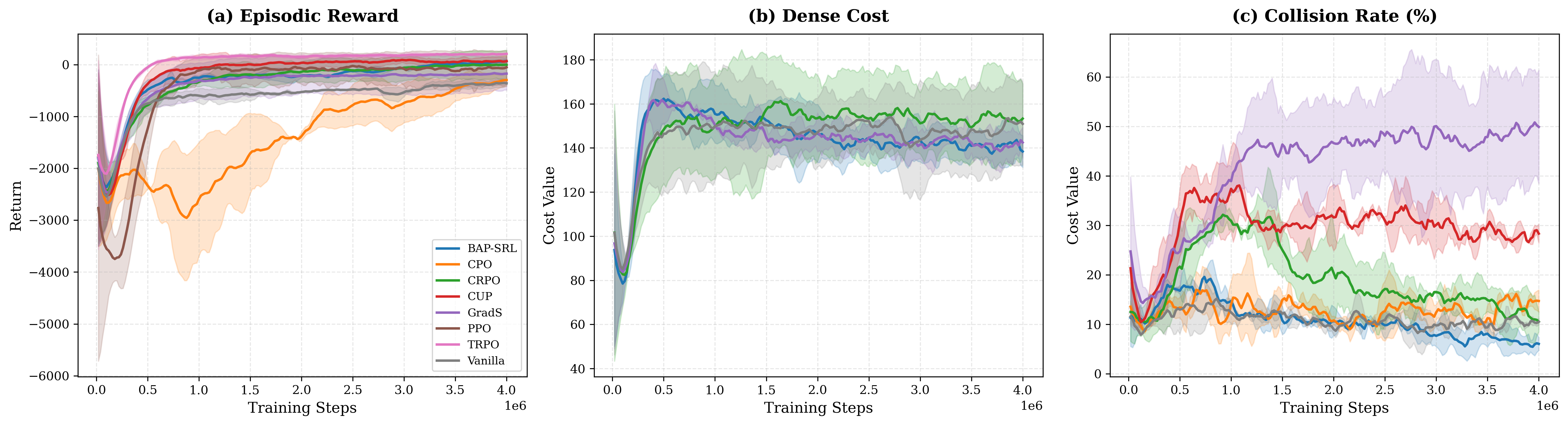}}%
    \caption{\textbf{Learning curves during training.} The solid lines represent the mean performance, and the shaded areas represent the standard deviation.}
    \label{fig:learning_curves}
\end{figure*}

\begin{table*}[t]
\centering
\footnotesize
\caption{Comparative Evaluation of Motion Planning Algorithms in Mixed Traffic Intersections}
\label{tab:main_results}
\begin{tabular}{l|l|cc|cc|c}
\toprule
\multicolumn{2}{c|}{} & \multicolumn{2}{c|}{\textbf{Safety}} & \multicolumn{2}{c|}{\textbf{Efficiency}} & \multicolumn{1}{c}{\textbf{Comfort}} \\
\cmidrule(lr){3-4} \cmidrule(lr){5-6} \cmidrule(lr){7-7}
\textbf{Category} & \textbf{Algorithm} & \textbf{Avg. Risk} $\downarrow$ & \textbf{Collision Rate} $\downarrow$ & \textbf{Avg. Speed} $\uparrow$ & \textbf{Time to Goal} $\downarrow$ & \textbf{Avg. Jerk} $\downarrow$ \\
& & & (\%) & (km/h) & (s) & ($m/s^3$) \\
\midrule
\multirow{3}{*}{Unconstrained} 
& TRPO \cite{schulman2015trust} & $0.56 \pm 0.21$ & $30.50 \pm 46.04$ & $14.11 \pm 2.13$ & $14.84 \pm 1.18$ & $4.02 \pm 0.41$ \\
& PPO \cite{schulman2017proximal} & $0.60 \pm 0.17$ & $44.00 \pm 49.64$ & \textbf{\boldmath $14.22 \pm 2.08$} & \textbf{\boldmath $14.68 \pm 1.15$} & $4.00 \pm 0.49$ \\
\midrule
\multirow{6}{*}{Constrained} 
& CPO \cite{achiam2017constrained} & $0.57 \pm 0.17$ & $34.29 \pm 47.47$ & $10.95 \pm 4.22$ & $17.59 \pm 1.61$ & \textbf{\boldmath $2.55 \pm 0.51$} \\
& CUP \cite{yang2022constrained} & $0.63 \pm 0.19$ & $34.34 \pm 47.49$ & $13.39 \pm 1.89$ & $15.73 \pm 1.76$ & $4.19 \pm 0.39$ \\
& PPOLag (Vanilla) \cite{ray2019benchmarking} & $0.53 \pm 0.16$ & $23.00 \pm 42.08$ & $11.16 \pm 3.28$ & $17.85 \pm 2.26$ & $3.29 \pm 0.61$ \\
& PPOLag (Min-Max) & $0.57 \pm 0.15$ & $26.26 \pm 44.01$ & $11.10 \pm 3.96$ & $17.36 \pm 1.95$ & $3.35 \pm 0.47$ \\
& PPOLag (CRPO) \cite{xu2021crpo} & $0.51 \pm 0.15$ & $16.50 \pm 37.12$ & $12.82 \pm 2.55$ & $16.72 \pm 2.12$ & $3.33 \pm 0.41$ \\
& PPOLag (GradS) \cite{yao2024gradient} & $0.63 \pm 0.22$ & $46.00 \pm 49.84$ & $12.75 \pm 3.22$ & $15.03 \pm 1.15$ & $3.92 \pm 0.54$ \\
\midrule
& \textbf{BAP-SRL} & \textbf{\boldmath $0.48 \pm 0.17$} & \textbf{\boldmath $9.00 \pm 28.62$} & \textbf{$13.27 \pm 1.49$} & \textbf{$16.88 \pm 1.18$} & \textbf{$4.66 \pm 0.35$} \\
\bottomrule
\end{tabular}%
\end{table*}

\subsection{Baseline Comparison}
\label{subsec:baseline_comparison}

We compare the proposed BAP-SRL against state-of-the-art unconstrained and constrained RL baselines. Fig. \ref{fig:learning_curves} presents the comparative learning curves over $4 \times 10^6$ training steps. In terms of task performance (Fig. \ref{fig:learning_curves}a), BAP-SRL demonstrates strong sample efficiency, converging to high rewards comparable to the unconstrained PPO baseline, whereas CPO suffers from significant performance degradation. Regarding safety performance (Fig. \ref{fig:learning_curves}b and Fig. \ref{fig:learning_curves}c), our method maintains stable dense costs and collision rate comparable to the baselines. While GradS reduces dense costs but fails to effectively minimize collision rate, BAP-SRL achieves the lowest final dense cost and collision rate, surpassing PPO-Lagrangian baselines in the late training stages. This validates that the Bayesian adaptive priority mechanism effectively balances reward seeking with safety compliance in complex intersection scenarios.

As shown in Table \ref{tab:main_results}, unconstrained methods (TRPO and PPO) achieve high efficiency but suffer from unacceptable safety violations, with collision rates exceeding 30\%. Among constrained methods, BAP-SRL demonstrates superior safety performance. It achieves the lowest collision rate of 9.00\% and the lowest average risk of 0.48. Notably, this represents a {60.87\%} reduction in collision rate compared to PPOLag (Vanilla) and a 73.75\% reduction compared to CPO. Compared to other Lagrangian methods, BAP-SRL also achieves a satisfactory efficiency performance, achieving a 19.55\% and 2.76\% improvement in average speed and time to goal compared to PPOLag (Min-Max). While BAP-SRL incurs an increase in average jerk ($4.66 m/s^3$), this reflects the necessary dynamic braking maneuvers required to resolve critical conflicts in dilemma zones.

Table \ref{tab:safety_breakdown} breakdowns the collision rate by hazard source. BAP-SRL drastically reduces collisions with VRUs to 2.00\%, a notable improvement over other baselines. Although rear-end and side collisions slightly increase to 3.00\% and 4.00\%, this remains within a manageable range compared to the significant gains in frontal collision avoidance.

Table \ref{tab:maneuver_robustness_sr} illustrates the generalization capability of BAP-SRL across different tasks. While baselines often struggle with specific tasks (e.g., PPOLag CRPO drops to $73.02\%$ success in right turns), BAP-SRL maintains high performance across all scenarios. It achieves the highest success rates in right turn (91.67\%) and straight ({90.00\%) maneuvers, and remains highly competitive in left turns ($91.67\%$), demonstrating stable conflict resolution regardless of task requirements.

\begin{table}[t]
\centering
\scriptsize
\caption{Collision rate (\%) breakdown by hazard source}
\label{tab:safety_breakdown}
\begin{tabular}{lccc}
\toprule
\textbf{Method} & \textbf{VRU} & \textbf{Rear Vehicle} & \textbf{Side Vehicle} \\
\midrule
TRPO \cite{schulman2015trust} & $22.50 \pm 41.76$ & \textbf{\boldmath $0.00 \pm 0.00$} & $8.00 \pm 27.13$ \\
PPO \cite{schulman2017proximal} & $36.00 \pm 48.00$ & \textbf{\boldmath $0.00 \pm 0.00$} & $8.00 \pm 27.13$ \\
CPO \cite{achiam2017constrained} & $21.43 \pm 41.03$ & $4.29 \pm 20.25$ & $8.57 \pm 27.99$ \\
PPOLag (Vanilla) \cite{ray2019benchmarking} & $16.00 \pm 36.66$ & $6.00 \pm 23.75$ & \textbf{\boldmath $1.00 \pm 9.95$} \\
PPOLag (Min-Max) & $11.11 \pm 31.43$ & $2.02 \pm 14.07$ & $13.13 \pm 33.77$ \\
PPOLag (CRPO) \cite{xu2021crpo} & $9.00 \pm 28.62$ & $1.00 \pm 9.95$ & $6.50 \pm 24.65$ \\
PPOLag (GradS) \cite{yao2024gradient} & $30.00 \pm 45.83$ & \textbf{\boldmath$0.00 \pm 0.00$} & $16.00 \pm 36.66$ \\ \midrule
\textbf{BAP-SRL} & \textbf{\boldmath $2.00 \pm 14.00$} & \textbf{$3.00 \pm 17.06$} & \textbf{$4.00 \pm 19.60$} \\
\bottomrule
\end{tabular}
\end{table}

\begin{table}[htbp]
\centering
\scriptsize
\caption{Success rate (\%) across intersection maneuvers}
\label{tab:maneuver_robustness_sr}
\begin{tabular}{lccc}
\toprule
\textbf{Method} & \textbf{Left Turn} & \textbf{Right Turn} & \textbf{Straight} \\
\midrule
TRPO \cite{schulman2015trust} & $72.13 \pm 44.84$ & $70.97 \pm 45.39$  & $66.23 \pm 47.29$ \\
PPO \cite{schulman2017proximal} & $48.65 \pm 49.98$ & $45.71 \pm 49.82$ & $78.57 \pm 41.03$ \\
CPO \cite{achiam2017constrained} & $31.25 \pm 46.35$  & $63.64 \pm 48.10$ & $84.38 \pm 36.31$ \\
PPOLag (Vanilla) \cite{ray2019benchmarking} & \textbf{\boldmath $92.31 \pm 26.65$}  & $65.79 \pm 47.44$ & $77.78 \pm 41.57$  \\
PPOLag (Min-Max) & $86.49 \pm 34.19$ & $55.88 \pm 49.65$ & $78.57 \pm 41.03$  \\
PPOLag (CRPO) \cite{xu2021crpo} & $89.33 \pm 30.87$ & $73.02 \pm 44.39$ & $87.10 \pm 33.52$ \\
PPOLag (GradS) \cite{yao2024gradient}  & $50.00 \pm 50.00$ & $48.28 \pm 49.97$ & $64.52 \pm 47.85$ \\ \midrule
\textbf{BAP-SRL} & \textbf{$91.67 \pm 27.64$} & \textbf{\boldmath $91.67 \pm 27.64$} & \textbf{\boldmath $90.00 \pm 30.00$}  \\
\bottomrule
\end{tabular}
\end{table}


\subsection{Ablation Study}
\label{sec:ablation}

\subsubsection{Bayesian Components}
The BAP mechanism fuses proior ($\Phi_{\rm prior}$) and instantaneous risk evidence ($\Phi_{\rm obs}$). To isolate the contribution of each term, we evaluate two variants:
\begin{itemize}
    \item {w/o Prior ($\alpha=0, \rho_k=0$):} The attention weight depends solely on the instantaneous violation $\Delta_k$, transforming into a purely reactive weight controller without memory of prior constraint difficulty
    \item {w/o Likelihood ($\beta=0$):} Lagrangian weights are determined only by the prior, ignoring real-time risk
\end{itemize}


Results in Table \ref{tab:ablation_components} reveal distinct failure modes in the ablated variants. 
Removing the prior leads to the highest collision rate ($23.00\%$), as a purely reactive approach fails to account for long-term constraint difficulty.
Similarly, the baseline without likelihood evidence results in the lowest average reward ($20.18$). This indicates that relying solely on priors might lead to conservative policies, as the agent becomes cautious to exploit safe gaps, failing to balance safety with operational efficiency.
The full BAP-SRL achieves the optimal balance, reducing the collision rate to 9.00\% while achieving a reward of 68.32, validating the effectiveness of our proposed framework.

\begin{table}[t]
\centering
\scriptsize
\caption{Ablation Analysis of Bayesian Components}
\label{tab:ablation_components}
\begin{tabular}{cccc}
\toprule
\textbf{Prior} & \textbf{Likelihood} & \textbf{Collision Rate (\%)} & \textbf{Avg. Reward} \\
\midrule
$\times$ & $\checkmark$ & $23.00 \pm 42.08$ & 56.72 \\
 $\checkmark$ & $\times$ & $18.00 \pm 38.42$ & 20.18 \\
 $\checkmark$ & $\checkmark$ & \textbf{\boldmath $9.00 \pm 28.62$} & \textbf{\boldmath $68.32$} \\
\bottomrule
\end{tabular}
\end{table}

\subsubsection{Static Priority}
A unique feature of BAP-SRL is the integration of semantic priors ($\rho_k$) to enforce safety preferences. We compare this setting against a non-biased version where all $\rho_k = \min\{\rho_{\rm VRU}, \rho_{\rm veh,side}, \rho_{\rm veh,rear}\} = -2.0$.

\begin{figure}[t!]
\centering
\includegraphics[width=\columnwidth]{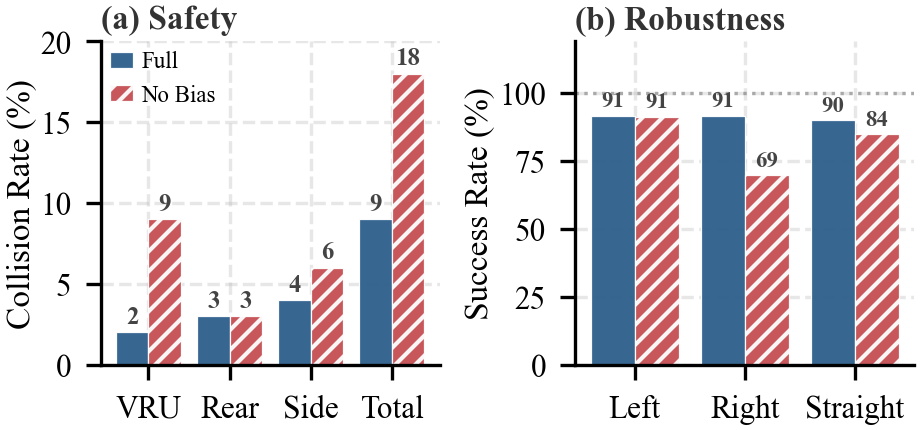}
\caption{\textbf{Ablation analysis of static priority.} (a) \textbf{Collision rate breakdown by object type.} The unbiased baseline results in a doubled total collision rate and significantly higher risk to VRUs. (b) \textbf{Success rate across different maneuvers.} The proposed BAP-SRL maintains high robustness, whereas the baseline performance drops drastically in complex right-turn scenarios.}
\label{fig:ablation}
\end{figure}

As illustrated in Fig. \ref{fig:ablation}, the absence of static priority leads to a significant degradation in safety performance. The unbiased baseline exhibits a doubled total collision rate ($18.00\%$ vs. $9.00\%$). Crucially, the collision rate with VRUs increases from $2.00\%$ to $9.00\%$. Furthermore, robustness in complex scenarios is severely compromised. The success rate for unprotected right turns drops drastically to $69.70\%$ compared to $91.67\%$ in the full BAP model. This suggests that static priors effectively break symmetry in multi-objective conflicts, guiding the policy toward safer trade-offs and resulting in a higher average reward.


\subsection{Case Study}
\label{sec:case_study}

To qualitatively validate the efficacy of BAP-SRL in resolving multi-objective conflicts, we conduct two representative case studies involving the ego vehicle making a left turn and going straight. In these scenarios, the ego vehicle encounters a dilemma zone characterized by multiple conflicting threats: an aggressive two-wheeler cutting across the intended path under the hesitant mode (Section \ref{sec:asd}), a tailgating rear vehicle, and a side vehicle that restricts lateral maneuverability.

Case study 1 involves the ego vehicle making a left turn.
The decision dynamics is analyzed in Fig. \ref{fig:case_study_mechanisms}, comparing our approach with the PPOLag (Vanilla) \cite{ray2019benchmarking}.
Notably, the initial myopic behavior of baseline agent forces a heavy deceleration when the VRU threat emerges, leading to a sharp increase in VRU safety cost. In contrast, the BAP-SRL agent in Fig. \ref{fig:case_mech_bap} demonstrates a strategic velocity adjustment, contributing to a decrease in VRU safety cost.

In the more critical scenario depicted in Fig. \ref{fig:case_study_mechanisms2}, the baseline agent fails to resolve the conflict, resulting in a collision with the VRU (Fig. \ref{fig:case_mech_vanilla2}). Fig. \ref{fig:case_mech_bap2} illustrates how BAP-SRL successfully manages simultaneous multi-source risks. The agent accelerates cautiously and executes a mild deceleration to yield to the VRU without violating the rear and side safety thresholds. This confirms that the BAP mechanism effectively transforms complex physical constraints into proactive driving behaviors, ensuring safety even in high-risk dilemma situations.

\begin{figure*}[t!]

    \centering
    \subfloat[PPOLag (Vanilla)]{
        \includegraphics[width=0.45\textwidth]{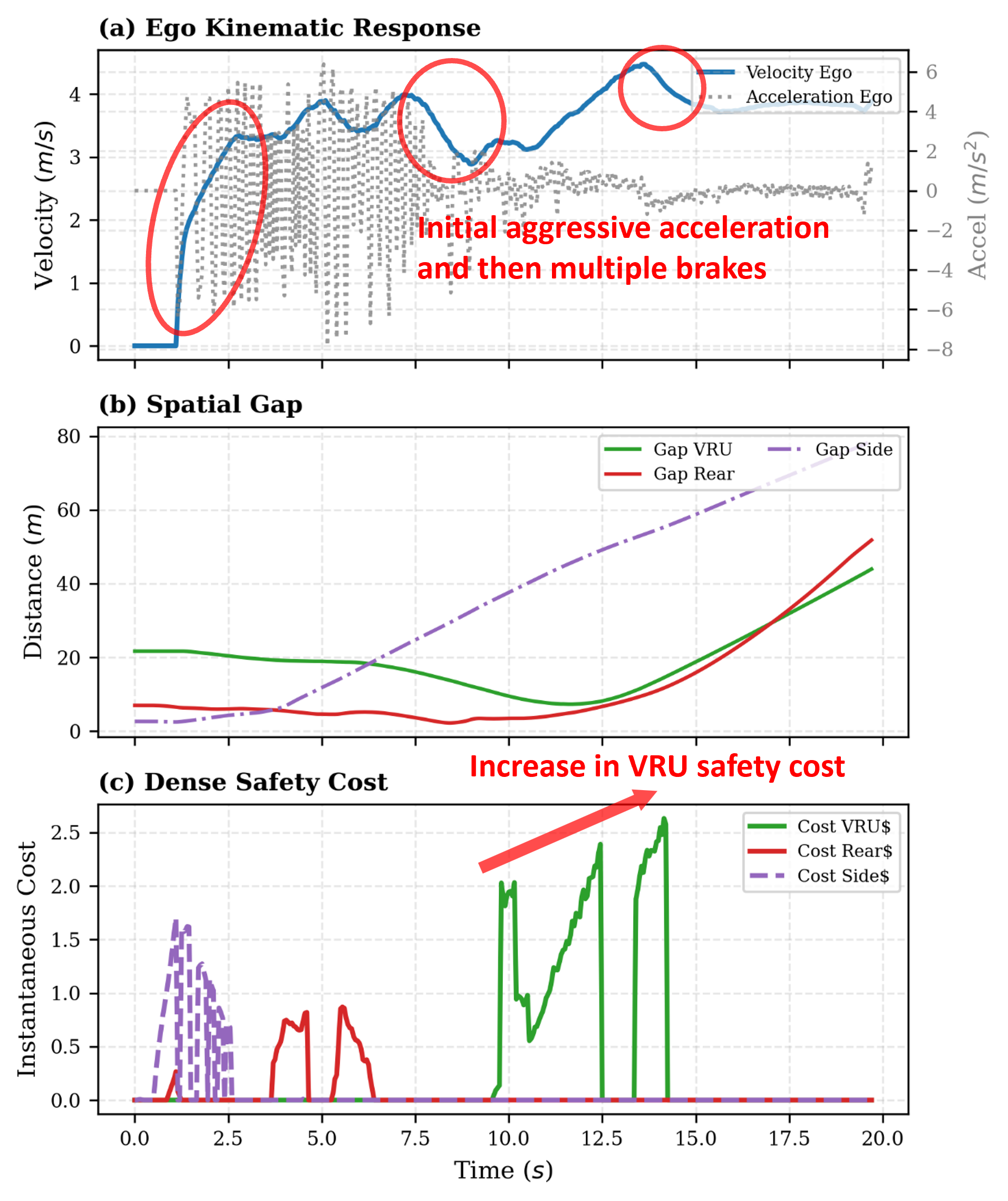}
        \label{fig:case_mech_vanilla}
    }  
    \subfloat[BAP-SRL]{
        \includegraphics[width=0.45\textwidth]{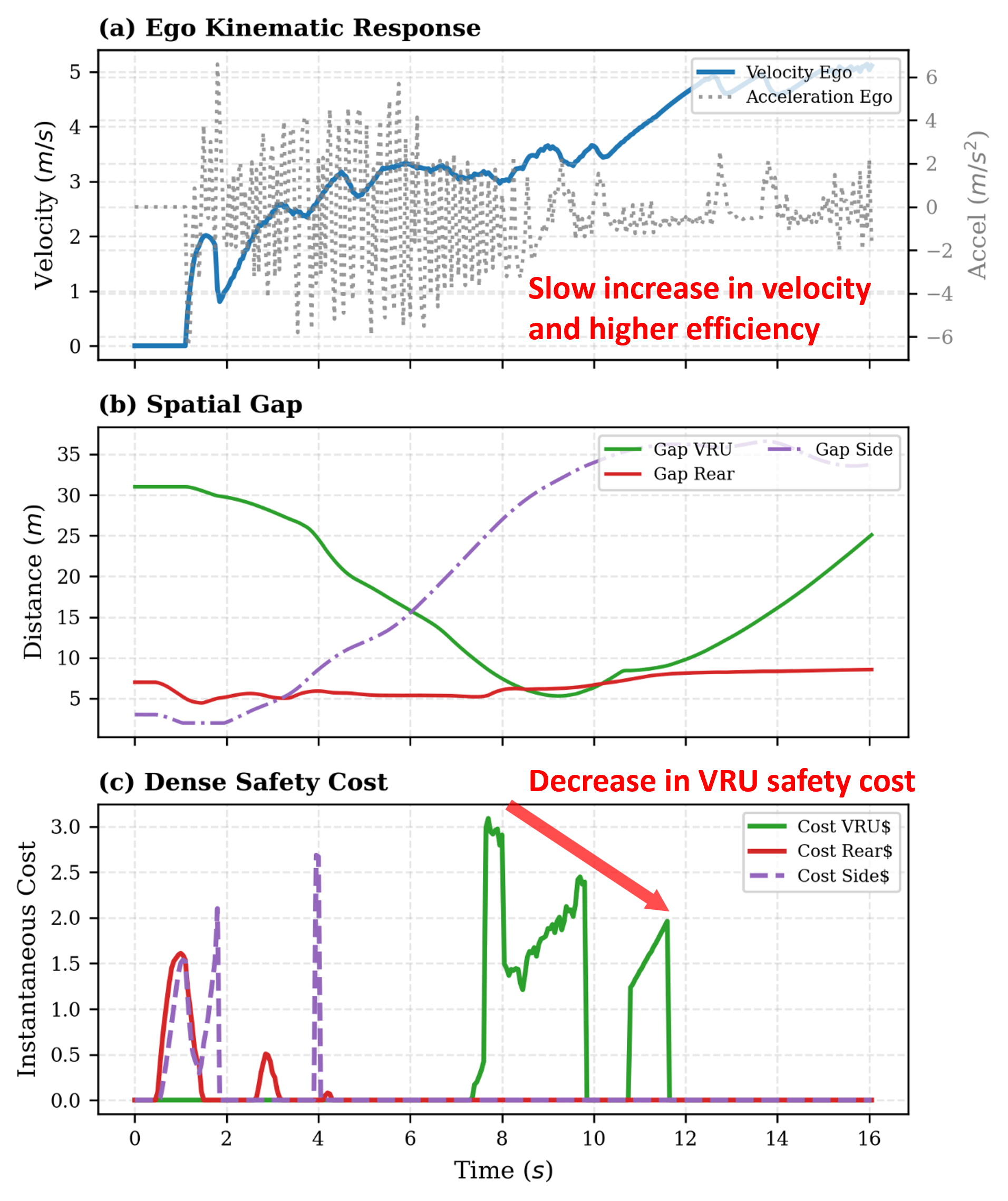}
        \label{fig:case_mech_bap}
    }
    \caption{\textbf{Comparative analysis of decision dynamics in sase study 1.} 
    The three subplots from top to bottom illustrate: (1) the ego vehicle's kinematic response (velocity and acceleration profiles), (2) spatial gaps between the ego and surrounding agents, and (3) dense safety costs. \textbf{(a) PPOLag (Vanilla):} The ego vehicle exhibits aggressive initial acceleration. Upon encountering the frontal VRU threat, it performs heavy deceleration, leading to an increasing VRU safety cost. 
    \textbf{(b) BAP-SRL:} The ego vehicle adopts a cautious initial acceleration to mitigate potential conflicts with rear and side vehicles. When approaching the VRU, it performs a probing behavior by maintaining a medium velocity. This strategy results in a decrease in VRU safety cost and completion time.
}
    \label{fig:case_study_mechanisms}
\end{figure*}

\begin{figure*}[t!]
    \centering
    \subfloat[PPOLag (Vanilla)]{
        \includegraphics[width=0.45\textwidth]{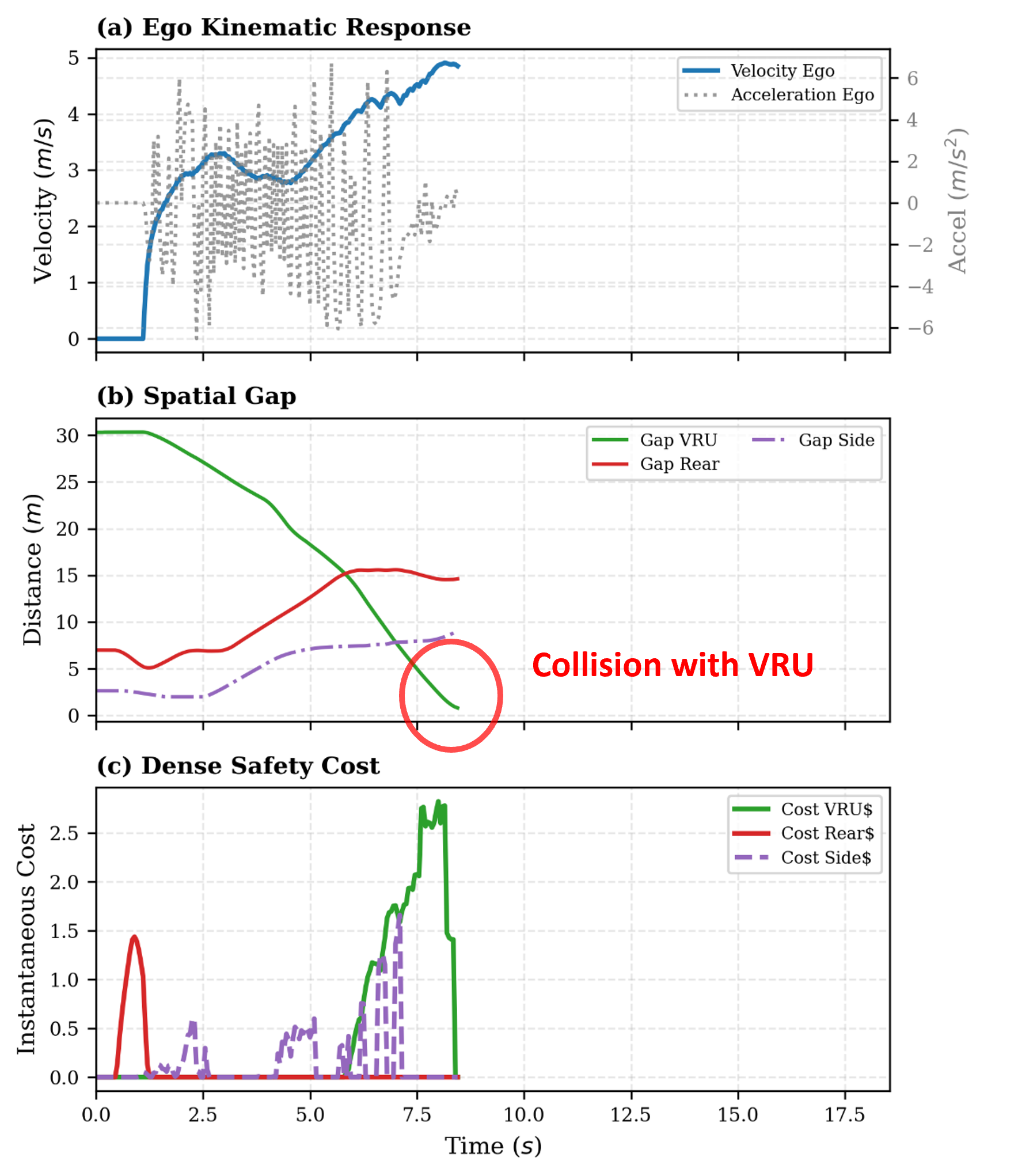}
        \label{fig:case_mech_vanilla2}
    }  
    \subfloat[BAP-SRL]{
        \includegraphics[width=0.45\textwidth]{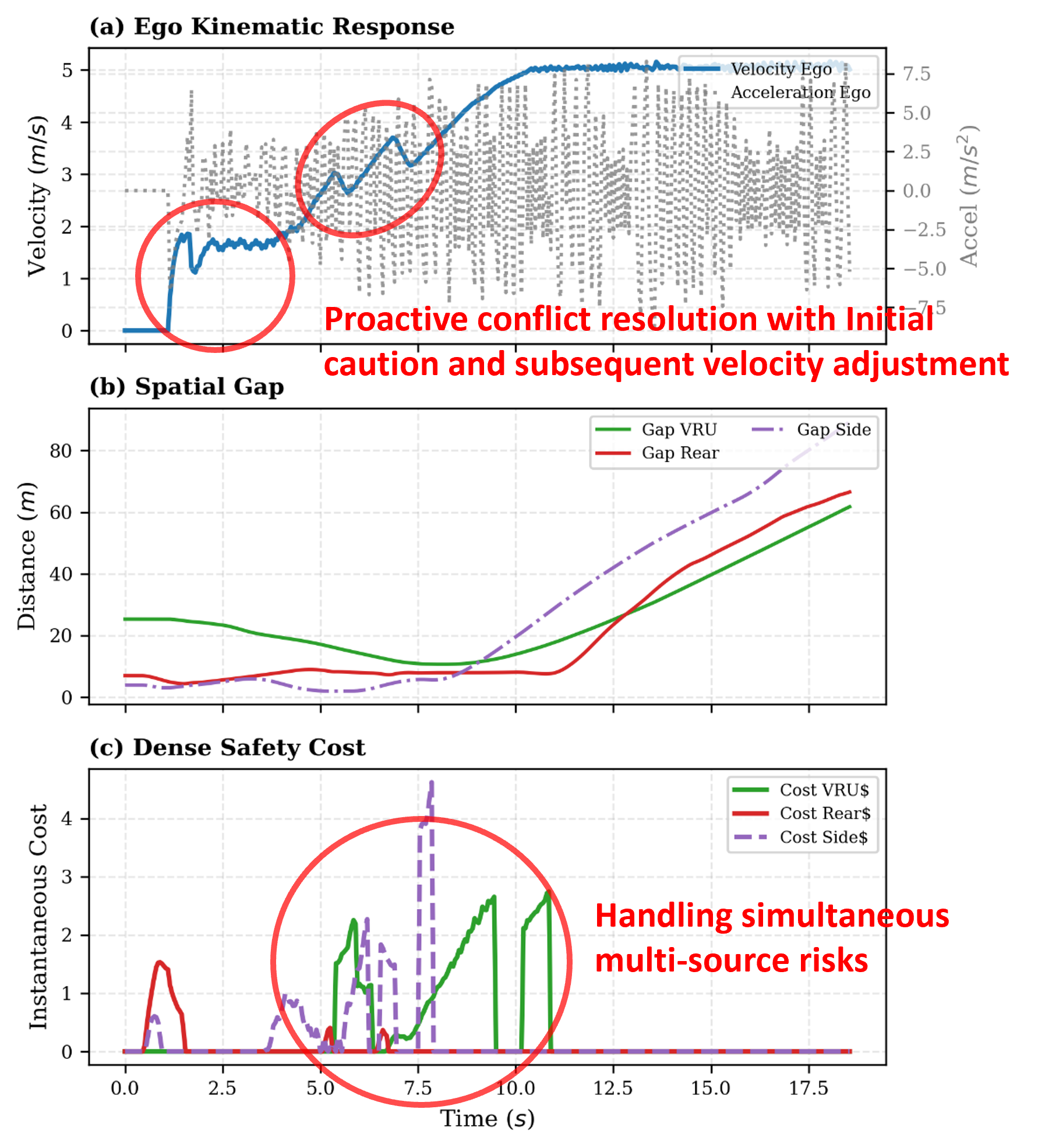}
        \label{fig:case_mech_bap2}
    }
    
    \caption{\textbf{Comparative analysis of decision dynamics in case study 2.} 
    The three subplots from top to bottom illustrate: (1) the ego vehicle's kinematic response (velocity and acceleration profiles), (2) spatial gaps between the ego and surrounding agents, and (3) dense safety costs. 
    \textbf{(a) PPOLag (Vanilla):} The ego vehicle ends up colliding with the VRU.
    \textbf{(b) BAP-SRL:} The ego vehicle handle multi-source risk through a cautious acceleration to assess the side threat and a controlled deceleration to yield to the VRU.}
    \label{fig:case_study_mechanisms2}
\end{figure*}

\section{Conclusions}

This paper presented BAP-SRL, a novel safe RL framework for safe motion planning at mixed traffic intersections. 
By formulating a Bayesian adaptivepriority mechanism, our approach effectively bridges the gap between long-term safety preferences priors and instantaneous risk evidence.
This allows the agent to maintain high task efficiency in safe conditions while instantaneously shifting focus to collision avoidance during critical states. Extensive experiments in CARLA demonstrate that BAP-SRL outperforms SOTA baselines. The results confirm that our approach effectively harmonizes the trade-off between efficiency and safety, providing a robust and scalable solution for safe motion planning in mixed traffic.

Despite these promising results, several limitations warrant further investigation. The current framework relies on ground-truth state information from the simulator, neglecting the perception noise and occlusion common in real-world scenarios. Future work is encouraged to incorporate uncertainty quantification into the Bayesian mechanism to enhance robustness against perceptual ambiguity.

\section*{Acknowledgement(s)}
This research is supported by grants from National Natural Science Foundation of China (No. 52325209, 52272420), Tsinghua University-Mercedes Benz Joint Institute for Sustainable Mobility, and Tsinghua-Toyota Joint Research Institute Inter-disciplinary Program.

\bibliographystyle{IEEEtran}
\bibliography{ref}

\vspace{11pt}

\vfill

\end{document}